\documentstyle[psfig,epsfig,twocolumn,openbib,cite]{article}
\newcommand{\xx}{{\bf x}}
\newcommand{\yy}{{\bf y}}
\renewcommand{\ss}{{\bf s}}
\newcommand{\Ss}{{\ss}}
\newcommand{\Eepsilon}{{\mbox{\boldmath $\epsilon$}}}
\newcommand{\Eeta}{{\mbox{\boldmath $\eta$}}}

\newcommand{\ff}{{\bf f}}

\newcommand{\be}{\begin{equation}}
\newcommand{\ee}{\end{equation}}
\newcommand{\bes}{\begin{eqnarray}}
\newcommand{\ees}{\end{eqnarray}}

\newcommand{\av}[1]{\langle #1 \rangle}

\newcommand{\tisean}{TISEAN }

\begin{document} 
\parskip 0pt
\title{The human ECG -- nonlinear deterministic versus stochastic aspects}

\author{Holger Kantz%
\thanks{Max Planck Institute for Physics of Complex
   Systems, N\"othnitzer Str. 38, D--01187 Dresden}
~and 
Thomas Schreiber%
\thanks{Physics Department, University of Wuppertal, D--42097 Wuppertal}}
\date{\small\today\\
\begin{quote}
  We discuss aspects of randomness and of determinism in electrocardiographic
  signals. In particular, we take a critical look at attempts to apply methods
  of nonlinear time series analysis derived from the theory of deterministic
  dynamical systems. We will argue that
  deterministic chaos is not a likely explanation for the short time
  variablity of the
  inter-beat interval times, except for certain pathologies. 
  Conversely,
  densely sampled full ECG recordings possess properties typical of
  deterministic signals. In the latter case, methods of deterministic
  nonlinear time series analysis can yield new
  insights.
\end{quote}}
\maketitle

\renewcommand{\Large}{\large}
\section{Introduction}
The physiology of the human heart is well understood 
mechanistically. Clear relations between the heart rate and many
regulatory influences are known. The shape of a single cycle ECG is
understood in detail and can be related to the different actions of the
heart during one cycle. On the other hand, electrocardiographic
recordings pose intricate problems of time series analysis due to the
different nature of the dynamical processes on different time
scales. The long-term evolution has been reported~\cite{oneoverf} to
show non-stationarity in the sense of power law correlations. At time
scales of a few hundred beats, non-stationarity becomes less
pronounced and the dominant feature are rapid, hardly predictable
low-amplitude fluctuations in the instantaneous heart rate which is
usually defined through inter-beat intervals.  At the time scale of a
heart beat, high resolution ECG signals show repetitive structures
specific of the beating mechanism.  A number of attempts have been
made to analyse various aspects of cardiac time series in the context
of nonlinear deterministic dynamical systems.

Deterministic chaos offers an interesting explanation for the emergence of
aperiodicity and unpredictability. Since rather simple systems can exhibit
chaos, one is lead to use nonlinear time series methods to verify whether such
a source of unpredictability is underlying a given observation.  In fact, the
concept of deterministic low-dimensional chaos has proven to be fruitful in the
understanding of many complex phenomena despite the fact that very few natural
systems have actually been found to be low-dimensional deterministic in the
sense of the theory.  In favourable situations this approach allows one to
determine the number of active variables, the stability properties of the
system with respect to perturbations, and even the equations of motion,
including applications like noise reduction or control.

Deterministic chaos is not the only, and not even the most probable source of
aperiodicity. The superposition of a large number of active degrees of freedom
can produce extremely complicated signals, which might not be distinguishable
from randomness. Stochasticity in the sense that a system is driven by
processes whose dynamics are too complex to be inferred from the information
stored in the observations is the most frequent source of unpredictability in
open systems and field measurements.  The fundamentals of deterministic chaos
as a theory are by now well established and described in a rich literature (for
example \cite{Ott}, see~\cite{KaplanGlass} for an introductory text). Nonlinear
time series analysis based on this theoretical paradigm is described in two
recent monographs~\cite{abarbook,KantzSchreiber}, and a number of conference
proceedings volumes~\cite{SFI,dyndis,freital}.

Several authors have applied nonlinear time series methods to ECG data, with
varying success. Many findings are accepted positively by cardiologists, but
have been contested by more mathematically oriented researchers. One of the
strongest arguments, the ability to apply chaos control
techniques\cite{rabbit}, has recently been put in question by the successful
control of a stochastic process~\cite{Collins}. From the understanding of
physiological mechanisms like the regulatory system and the conduction system,
one is lead to expect certain deterministic structures.  However, the average
heart rate of the healthy heart depends deterministically on several different
processes, which in turn depend on other influences, so that this dependency
altogether might be far too complicated to be observable as a deterministic
process through ECG data alone.  It is our point of view, and we will give some
non-exhaustive support, that ECG data does not prove heart rate variablity to
be the result of an autonomous deterministically chaotic system. We will
however demonstrate that even without determinism there are situations in which
nonlinear time series analysis yields new insights.

\section{Determinism from time series}
Determinism in the mathematical sense means that there exists an autonomous
dynamical system, defined
typically by a first order ordinary differential equation $\dot\xx=\ff(\xx)$ in
a state space $\Gamma\subset{\bf R}^D$, which we assume to be observed
through a single measurable quantity $h(\xx)$. 
The system thus possesses $D$ natural variables, but the measurement
is a usually nonlinear projection onto a scalar value (the following discussion
can be straightforwardly extended to multi-channel measurements). In order to
recover the deterministic properties of the system, we have to reconstruct an
equivalent of the subspace of $\Gamma$ which is explored by the solutions of
the system from the observations. The time delay embedding method is a way to
do so that can be derived with mathematical rigour~\cite{embed}.  From
the sequence of $N$ scalar observations $s_1,s_2,\ldots,s_N$, overlapping
vectors $\ss_n=(s_n,s_{n-\tau},\ldots,s_{n-(m-1)\tau})$ are formed ($\tau$ is a
delay time). For mathematically perfect, noise free observations $s_n$, it is
proven that for $m>2D_f$ there exists a one-to-one relation between
$\ss_n\in{\bf R}^m$ and the unobserved vectors $\xx_n$ in the state space, from
which the measurements were taken. Here $D_f$ is the box-counting dimension of
the attractor, that is, of the set in state space which is visited
asymptotically by a trajectory. 

The standard approach towards establishing low-dimensional determinism in a
scalar time series consists of constructing spaces with increasingly larger
$m$ and to search for deterministic structures in each of these spaces. A
common tool is the correlation dimension introduced by Grassberger
\& Procaccia \cite{GP}, which tests for self-similarity of the set of points
and thus for the existence of a finite dimensional attractor. Despite the
existence of several pitfalls, this is still the method of choice when a finite
dimension is to be established. Since contamination by about 2\% of noise
usually suffices to destroy all nontrivial self-similarity, dimension estimates
{\sl per se} cannot be recommended in field measurements, even if they stem
from a deterministic source plus observational noise.  A more promising
approach is to establish the existence of nontrivial dynamical
correlations by either directly constructing equations of motion (forecast
maps) or by proving an enhanced predictability, when the dimensionality of the
embedding space is sufficiently high.

\section{Inter-beat intervals}
In this article we discuss exclusively electrocardiographic (ECG)
data~\cite{MIT} measured with surface electrodes. Some arguments can be carried
over to invasive recordings as well.  The ECG signal reflects the
electro-chemical activity of the cardiac muscle fibres, integrated over the
whole surface with a local weight factor, depending on the relative distance
and electrical conductivity between each point and the electrode.

In studies of the heart rate variablity, often the long-term ECG recording is
reduced to the sequence of inter-beat time intervals (RR-intervals), as defined
by the time between consecutive R-waves. (See Goldberger and
Goldberger~\cite{goldbook} for an introduction to clinical electrocardiography
and definitions of the nomenclature.)  However, there is a fundamental
difference between the full ECG trace and the sequence of inter-beat
intervals. In the latter, the ``time'' index is given by the event number and
the observable by the time elapsing in between two events.  This time is not
obviously a phase space observable in the sense of embedding theory, but recent
work~\cite{Sauer_t,Hegger+} implies that in principle RR-intervals may be
suited to search for determinism in the ECG.  However, attractors reconstructed
from inter-bear time intervals of similar duration possess extremely small
diameters. Small errors or noise in the time intervals can easily extinguish
all nontrivial sub-structure.

If indications of deterministic structure are weak, one can first try to reject
certain null-hypotheses. One of the simplest is that the data stem from a
stationary linear Gaussian random process (GRP).  It is obviously too simple
and can be rejected for most interesting data sets, since the marginal
distributions are almost never Gaussian. The default explanation for this fact
are {\sl static nonlinearities}, that is, nonlinearities in the measurement
process, which have nothing to do with the dynamics, but may hide the original
character of the process. Time reflection invariance, which is another property
of GRPs, is not affected by static nonlinearities in the measurement. It can,
however, be affected by a memory effect of the measurement device, for example
by a simple causal low-pass filter on the data.

\begin{figure}
\centerline{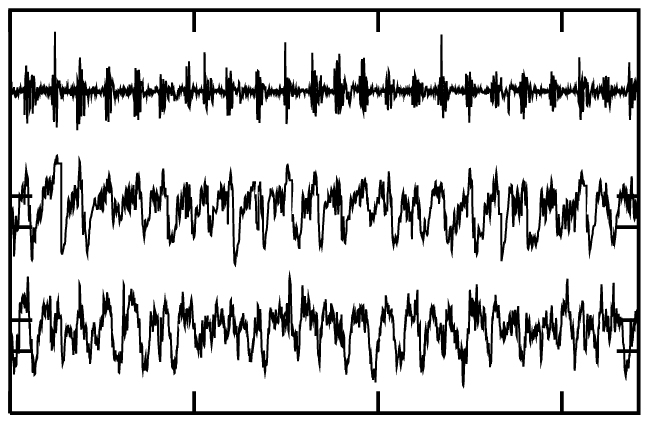} 
\caption[]{\label{heartrate}\small Simultaneous measurements of breath and
   heart rates~\cite{gold}, upper and middle trace. The lower trace contains a
   randomised sequence preserving the autocorrelation structure and the cross
   correlation to the fixed breath rate series. The surrogate data mimic much
   of the structure contained in the data, but not all of it.}
\end{figure}

In order to test against a composite null hypothesis, one can produce new data,
so called surrogates, and compare the original data and the surrogates by help
of a suitably chosen test statistics. The advantage is that we do not have to
specify, for example, {\em which} particular process generated the data but we
can test if {\em any} GRP could be underlying.  For this approach, first an
ensemble of surrogate data sets has to be produced. If the null hypothesis is
that all structure of the data is fully described by their marginal
distribution and by their auto-correlation function (equivalently, by their
power spectrum), one creates otherwise random data with exactly the same power
spectrum and marginal distribution~\cite{theiler1,Thomas_constrained}.  This
simple test will often yield a rejection of the null and thus suggest the
existence of additional, nonlinear structure, but can also be caused by
non-{\em stationary} features which are common in  clinical time series
and which are not compatible with the null hypothesis.

Figure~\ref{heartrate} illustrates the variation of the heart rate (middle
trace) of a human at rest in connection with the breath rate (upper trace).
Here, the instantaneous heart rate was derived from the RR-interval series by a
Fourier interpolation technique, see~\cite{gold} for details.  The lower
trace contains a randomised sequence that has the same autocorrelation
structure and the same cross correlation to the fixed breath rate recording.
These random surrogate data preserve much but not all of the structure observed
in the measurements. The remaining structure, in particular a slight asymmetry
under time-reversal, might be due to the measurement process. In any case the
structure is not pronounced enough for any kind of deterministic modeling.

\begin{figure}
\centerline{\psfig{file=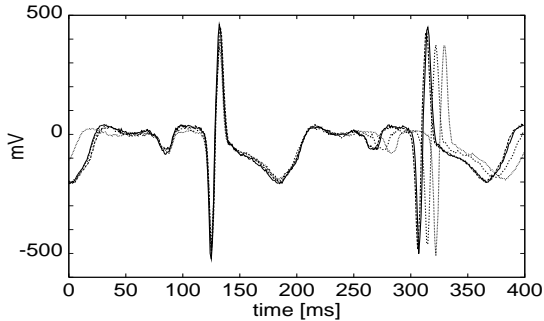,width=7.5cm,height=4.5cm}}~\\[-1cm]
\caption[]{\label{fig:superimpose}\small
   Superposition of four consecutive R-waves. The length of the baseline
   interval between T- and P-wave varies, whereas the cycles from P to T look
   almost identical.}
\end{figure}

It is hard to find nonlinearities in RR-interval time series beyond those which
can be easily accounted for by the measurement process or by the coupling to
the slower breath rate.  We thus support the conclusion of~\cite{nochaos}, that
there is no clear evidence for determinism in RR-interval data.  This does not
mean that determinism is positively absent, but that potential determinism in
the heart rate variability is too complex to be accessible through RR-interval
data alone. The conservative working hypothesis, which is also implied by the
widespread use of spectral characteristics, is that the process which governs
the initiation of new cardiac cycles is effectively stochastic, superimposed by
the regulations of the autonomous nervous system which control the average
heart rate. This does not in itself exclude the possibility that methods for
the characterisation of deterministic chaos (like dimension estimates,
entropies, empirical symbolic encoding) can be exploited for diagnostic
purposes. In the next section we will show why despite the supposed lack of
determinism certain nonlinear analysis tools can be employed, focusing on the
single cycle ECG wave.

\section{Embedding of single-cycle ECG waves}
The fact that the different parts of the single cycle ECG wave mirror well
understood physiological processes which are repeated with considerable
precision introduces determinism-like properties in the ECG signal during the
time between the beginning of the P-wave and the end of the T-wave. A
simplified model of the ECG signal could consist of a concatenation of
identical PQRST complexes, with a time interval of short but random duration in
between. This is motivated by Fig.~\ref{fig:superimpose}, where 4 consecutive
QRS-complexes from a long ECG recording are superimposed. Although in general
there is much more variability in the wave-form, Fig.~\ref{fig:superimpose}
shows impressively the reproducibility of the wave-form with random TP
intervals in between. A more detailed study which is in progress shows that the
whole variablity of a single cycle ECG wave can be parametrised by a few,
typically about 5, parameters.

\begin{figure}
\centerline{\hspace*{-1cm}\psfig{file=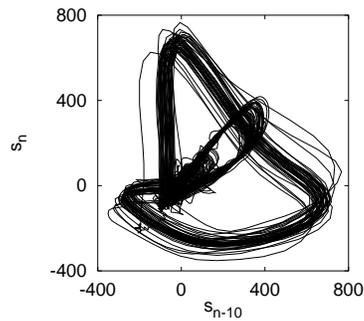,width=5cm}}~\\[-1cm]
\caption[]{\label{fig:embed}\small Two-dimensional embedding of ECG data
   covering a time of 20~s, plotted with a delay of 10~ms.}
%%silc_a.dat_c, delay 10, 20000 Datenpunkte
\end{figure} 

Single-cycle ECG-waves which can be well approximated by lower dimensional
surfaces. This fact allows to make use of methods of nonlinear time series
analysis which exploit the concept of phase space. In Fig.~\ref{fig:embed} we
show an ECG signal in a two-dimensional time delay embedding space. In this
representation, the data seem to fall onto a two-dimensional ribbon, but
including more delay coordinates one would find a higher dimensional
hyper-tube.  We do not say that the data lie on an ``attractor'', since the
object we see is not invariant in the strict sense: when accumulating more and
more data, due to non-stationarity, they will fill more and more of the
plane. Nevertheless, the heart rate variability and the unpredictability of the
onset of the next heart cycle in this plot is hidden in the cluster around the
origin, so that we see essentially the few-parameter set of single cycle
segments. This will allow to approximate this set locally by smooth manifolds.

\begin{figure}
\centerline{\psfig{file=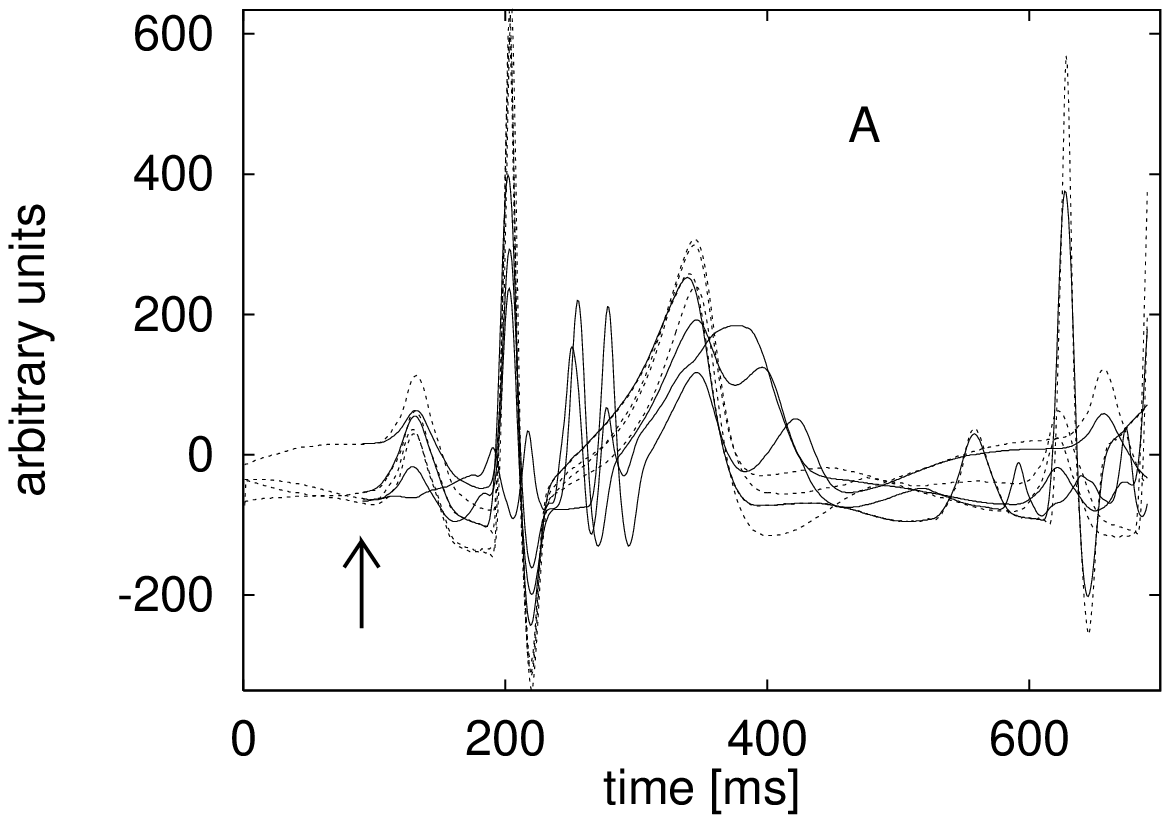,width=7cm,height=4cm}}
\centerline{\psfig{file=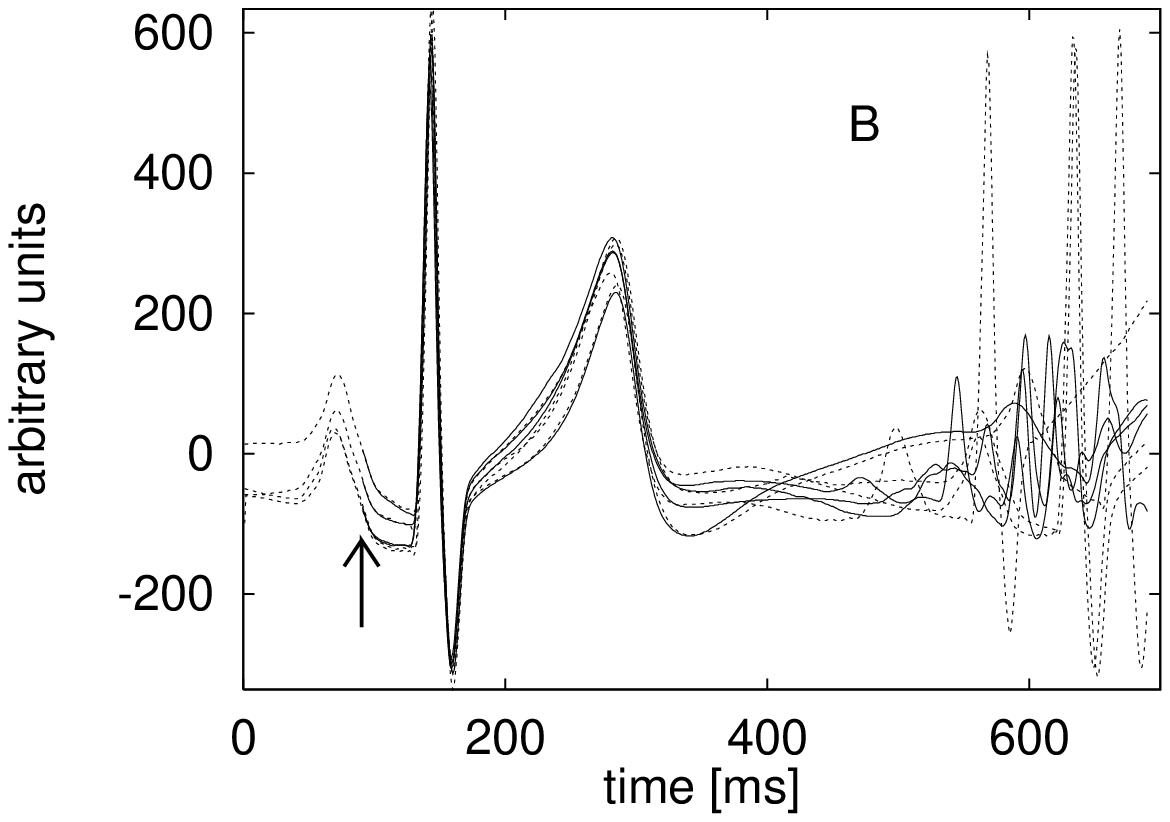,width=7cm,height=4cm}}
\caption[]{\label{fig:forecast}\small Phase space predictions 
   (continuous lines), compared to measurements (dotted lines): Using
   the last 50 measurements represented by the broken lines to the left of the
   arrow to characterise the present state, the future was predicted. In panel
   A, predictions start 160~ms before the R-wave, in B 100~ms.  In each panel,
   four different cycles are shown.}
\end{figure}

\begin{figure}
%\centerline{\psfig{file=errors.ps,width=7cm}}
\centerline{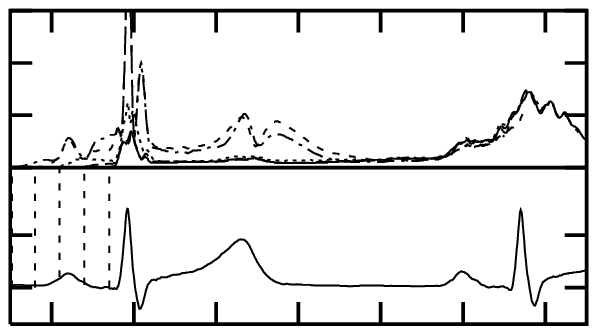}~\\[-1cm]
\caption[]{\label{fig:forecast_errors}\small Average forecast errors
   for predictions from 1 to 600 ms ahead.  Predictions start from segments
   $\Ss_n$ consisting of the 50 data items immediately before the dotted lines
   relating the position to the point in the ECG-cycle (reduced in size). All
   forecasts/data segments were synchronised by the zero crossing of the
   descending part of the R-wave.}
\end{figure}

{\em Predictability.} The existence of (approximate) constraints like a
containing manifold can be used to make short time predictions inside an
ECG-cycle. Strict determinism would mean that a point in state space possesses
a unique future, and that nearby points have a similar future. Thus, collecting
neighbours of the point whose future should be determined, we can average over
the future of the neighbours to learn how the present signal will
continue. This simplest implementation of phase space predictability exploits
much more than just linear temporal correlations.  The present state is
characterised by a delay vector of sufficient length, or, otherwise said, by a
segment of measurements. To obtain the predictions discussed below we select
such a 50~ms data segment, look for similar segments in the data base, and take
the average over the continuations of all these similar segments as a
prediction of the future of the this data segment. Let
$\Ss_n=(s_n,\ldots,s_{n-49})$ be the segment whose future should be determined,
and $(s_{n'},\ldots,s_{n'-49})$ a similar segment, i.e.
$|s_{n-i}-s_{n'-i}|<\delta$ for all $i=0,\ldots,49$, then the prediction $\hat
s_{n+k}$ $k$ ms ahead is $s_{n'+k}$ averaged over all similar segments $n'$.
In Fig.~\ref{fig:forecast} we show a few typical results of such predictions
(continuous lines) up to $k_{\rm max}$=600~ms ahead (to be compared to the
measurements, dotted), starting from two different locations on an ECG cycle.
In panel A, $\Ss_n$ is a 50~ms segment of data inside the interval between T-
and P-wave. In panel B, $\Ss_n$ contains a good part of the P-wave. In
Fig.~\ref{fig:forecast_errors} we show an average of the prediction errors, the
modulus of the difference between the predicted and the observed value, for
different starting positions. The averages were made over 306 cycles.

Let us emphasise three observations: (1) When making predictions for segments
which contain only data before the beginning of the P-wave (the two curves
starting leftmost in Fig.~\ref{fig:forecast_errors}) the predictions fluctuate
quite soon, leading to large errors even a few steps ahead.  Small fluctuations
of this part of the signal are mis-interpreted by the algorithm to indicate the
beginning of the next P-wave. On this time scale, ECG data do not contain much
information about the onset of the next heart beat. (2) As soon as the true
beginning of the P-wave is covered by $\Ss_n$ (the three other curves in
Fig.~\ref{fig:forecast_errors}), predictions are very robust and accurate: Even
the end of the T-wave (250 ms ahead) is predicted quite well, and the forecast
errors are almost zero up to 400~ms ahead, the beginning of the next P-wave.
(3) The predictions do not become worse with increasing prediction time (up to
the above mentioned 400~ms), as it would be for chaotic data.  Altogether,
these findings show that each cycle can be characterised by a single state
vector. The initial part of the P-wave already determines position, amplitude
and width of the T-wave. Of course, all this is in agreement with physiological
knowledge.

{\em Noise reduction.} 
Due to the sharp QRS-complexes, the power spectrum of an ECG signal is broad
banded. Therefore, filters in Fourier space  cannot be employed with great
success. They might smooth the signal on average, but also destroy the
structure of the QRS-complex. However, nonlinear techniques originally designed
for the treatment of noise in low-dimensional chaotic data can be applied, as
long as the approximation by a manifold introduces less errors than there are
measurement errors on the data.

Conceptual as well as technical issues of nonlinear noise reduction for chaotic
data have been well discussed in the literature. (See~\cite{ks} for a review
containing the relevant references.)  Phase space projection
techniques rely on the assumption that the signal of interest is
approximately described by a manifold that has a lower dimension than some
phase space it is embedded in, and only the noise contamination of the data
drives them out of this hypersurface.  Then noise is reduced by first
identifying the manifold {\sl in the noisy data} and then by projecting the
noisy data onto this manifold.  This can be formalised as follows. Let
$\{\xx_n\}$ be the states of the system at times $n=1,\ldots,N$, represented in
some space ${\cal R}^d$.  A $(d-Q)$-dimensional submanifold ${\cal F}$ of this
space can be specified by $F_q(\xx)=0,\quad q=1,\ldots,Q$. Even if the data are
not exactly constrained to ${\cal F}$, we can always find $\{\Eepsilon_n\}$
such that $\xx'_n=\xx_n+\Eepsilon_n$ and
$F_q(\xx_n+\Eepsilon_n)=0$ for all $q,n$.
Then $\sqrt{\av{\Eepsilon^2}}$ denotes the (root mean squared) average error we
make by approximating the points $\{\xx_n\}$ by the manifold ${\cal F}$. Assume
that in a measurement we obtain noisy data $\yy_n=\xx_n+\Eeta_n$, where
$\{\Eeta_n\}$ is some random contamination. By projecting these values onto the
manifold ${\cal F}$ we may find $\xx'_n=\xx_n+\Eepsilon_n$. If we can find a
suitable manifold and carry out the projections such that
$\av{\Eepsilon^2}<\av{\Eeta^2}$, then we have reduced the observational error
more than we distorted the data.

\begin{figure}
   \centerline{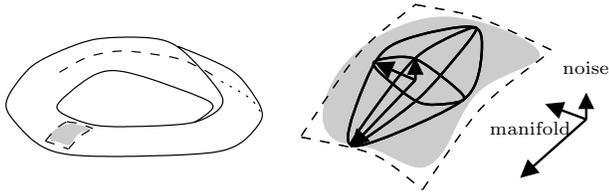} \caption[]{\small Illustration of the local
   projection scheme. For each point to be corrected, a neighbourhood is formed
   (grey shaded area) which is then approximated locally by an ellipsoid. An
   approximately two-dimensional manifold embedded in a three-dimensional space
   could for example be cleaned by projecting onto the first two principal
   directions.\label{fig:noise}}
\end{figure}

\begin{figure}
\centerline{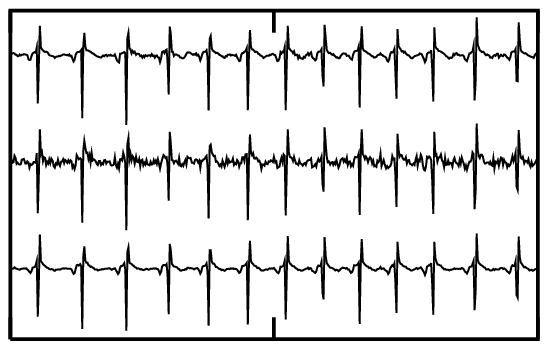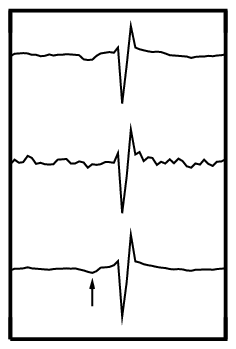}
\caption[]{\small Nonlinear noise reduction applied to electrocardiogram
  data. Upper trace: original recording. Middle: the same contaminated with
  typical baseline noise. Lower: the same after nonlinear noise reduction.  The
  enlargements on the right show that indeed clinically important features like
  the small downward deflection of the P-wave preceding the large QRS-complex
  are recovered by the procedure. Note that the noise and the signal have very
  similar spectral contents and could thus not be separated by Fourier
  methods.\label{fig:ecg}}
\end{figure}

The most practical way to approximate data by a manifold is by local linear
projections, which are obtained from local principal components. The procedure
is illustrated in Fig.~\ref{fig:noise}. The correction is done for each
embedding vector. Fig.~\ref{fig:ecg} shows the result of the noise reduction
scheme applied to a noisy ECG. For the purpose of noise reduction, embeddings
in higher dimensions are advantageous. Fig.~\ref{fig:ecg} was produced with
delay windows covering 200~ms, that is, $m=50$ at a delay time of 4~ms.
See~\cite{ecg} for more details on the nonlinear projective filtering of ECG
signals.

{\em Fetal ECG extraction.} Noise reduction can be regarded as the particular
case of the more general task of signal separation where one of the signals is
the noise contribution. It turns out that the methodology developed for noise
reduction can be generalised to the separation of other types of signals. Here,
we want to discuss the extraction of the fetal electrocardiogram (FECG) from
non-invasive maternal recordings. Other very similar applications include the
removal of ECG artefacts from electro-myogram (EMG) recordings (electric
potentials of muscle) and spike detection in electro-encephalogram (EEG) data.

\begin{figure}
\centerline{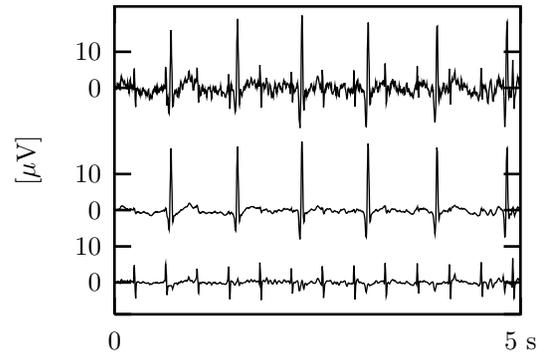} \caption[]{\small Signal separation by locally
  linear projections in phase space.  The original recording (upper trace)
  contains the fetal ECG hidden under noise and the large maternal
  signal. Projection onto the manifold formed by the maternal ECG (middle)
  yields fetus plus noise, another projection yields a fairly clean fetal ECG
  (lower trace). The data was kindly provided by
  J.~F. Hofmeister~\cite{recording}.\label{fig:fecg}}
\end{figure}

Fetal ECG extraction can be regarded as a three-way filtering problem since we
have to assume that a maternal abdominal ECG recording consists of three main
components, the maternal ECG, the fetal ECG, and exogenous noise, mostly from
action potentials of intervening muscle tissue. The two ECG components and the
noise have quite similar broad band power spectra and cannot be filtered apart
by spectral methods. In Ref.~\cite{fetal2}, it has been proposed to use a
nonlinear phase space projection technique for the separation of the fetal
signal from maternal and noise artefacts. A typical example of output of this
procedure is shown in Fig.~\ref{fig:fecg}. The assumption made about the
nature of the data is that the maternal signal is well approximated by a
low-dimensional manifold in delay reconstruction space. After projection onto
this manifold, the maternal signal is separated from the noisy fetal
component. Now it is assumed that the fetal ECG is also approximated by a
low-dimensional manifold and the noise is removed by projection. Since both
manifolds are curved, the projections have to be made onto linear
approximations. For technical details see Refs.~\cite{fetal2}.  The algorithms
for noise reduction and signal separation are publicly available as part of
the TISEAN project~\cite{tisean}.

\section{Non-stationarity}
\begin{figure}
%\centerline{\psfig{file=silc_b.stat.cps,width=5.5cm}}
\centerline{\psfig{file=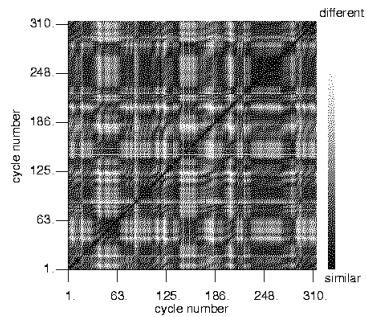,width=5.5cm}}
\caption[]{\label{fig:stationarity}\small Gray-scale coding of the pairwise
   $L^1$-distance between single-cycle wave trains as a function of the
   cycle numbers. A pixel at $(i,j)$ is encoded dark (black), if the wave
   trains $i$ and $j$ are similar (identical), and bright otherwise.}
\end{figure}

The last issue we want to discuss in this paper is the inevitable problem of
non-stationarity. As mentioned before, the ECG seems to show structures on all
time scales, formalised by the notion of $1/f$-noise. Most obviously, the heart
rate changes depending on the physical activity of an individual. On the other
hand, from a diagnostic point of view, important information is contained in
the variability which is summarised under the term non-stationarity.  It is
therefore important not to regard non-stationarity only as a complication that
arises for the statistical description of ECG data but as an important source
of information. So far, there is no general methodology of extracting the
relevant aspects from the vast amount of data collected in long-term
recordings.  A representation of the ECG in a (reconstructed) phase space may
offer a powerful basis to trace changes in the wave-form. Consider an increase
in the instantaneous heart rate.  The ECG cycle will appear more condensed
which is however not a simple rescaling described by one parameter. The
wave-form will occupy a different part of the time delay embedding
space. Searching for neighbours in this space, one will be able to identify
similar situations in a long recording and thus be able to subdivide it into
different phases~\cite{MS,Schreiber_nonstationary}. Due to the lack of chaos in
the ECG case, the cross-prediction errors used
in~\cite{Schreiber_nonstationary} can be replaced by phase space distances
between data segments covering the part of a cycle from the beginning of the
P-wave until the end of the T-wave. In Fig.~\ref{fig:stationarity} we show a
typical outcome of such a similarity study. The representation is similar to
the well-known recurrence plot~\cite{cas_recurr} where only recurrences of full
P-to-T segments are considered.

\section{Conclusions}
One or two decades ago, chaos theory was received with great enthusiasm: It
seemed to supply concepts for the analysis and understanding of aperiodic
temporal evolution in arbitrary environments. During the last years disillusion
prevailed, since it became clear that chaos in a stronger sense is hardly to be
expected outside well controlled laboratory experiments. Much effort has been
spent, or wasted, on the discussion ``is it chaos or is it noise''. In almost
all interesting dynamical problems in nature, stochastic fluctuations,
intrinsic instability, and a changing environment act together to produce the
intriguing patterns we observe. Given that the human heart is also likely to be
such a mixed system, we want to promote a different approach towards the issue
by asking what is a {\em useful} framework for the analysis of cardiac data.
Due to the presence of approximately deterministic structures in the cardiac
cycle, state space concepts can be applied with the necessary care when the
typical time scales of a cycle are involved in the analysis. On longer time
scales, the reduction of the available information to a single number, the
RR-interval is too severe to still permit a useful study with phase space
methods. When this reduction is avoided, comparisons in phase space may well be
suitable to study the variability of the cardiac dynamics over longer times.

We have argued that the heart dynamics contains a component which -- at least
on the basis of the information supplied by ECG recordings -- has to be
considered as stochastic. On the other hand, we have shown that the single
cycle ECG wave can be represented by only a few parameters, or, equivalently,
by state vectors in a low-dimensional embedding space. We presented two
applications, noise reduction and fetal ECG extraction. They demonstrate that
the concepts of nonlinear time series analysis are powerful also for
non-deterministic systems, if one is able to isolate a specific problem.
 
This work was supported by the SFB 237 of the Deutsche Forschungsgemeinschaft.

\bibindent=0cm

\end{document}